\documentclass[reprint, amsmath,amssymb, aps]{revtex4-2}
\usepackage{graphicx}
\usepackage{xcolor}
\usepackage{float}
\usepackage{dcolumn}
\usepackage{bm}
\usepackage{natbib}
\usepackage{hyperref}
\usepackage[bottom]{footmisc}

\begin{document}

\title{Temperature Fluctuations and quantum corrections near Black Hole Horizon}
\author{Anamika Avinash Pathak}
\email{p20190459@hyderabad.bits-pilani.ac.in}

\author{Swastik Bhattacharya}
\email{swastik@hyderabad.bits-pilani.ac.in}

\affiliation{Department of Physics, Birla Institute of Technology and Sciences-Pilani \\Hyderabad, 500078, India}

\date{\today} 

\begin{abstract}
Spatially varying near-horizon fluctuations of temperature of a Schwarzschild Black Hole is considered within the Euclidean Gravity approach. We present evidence that suggests that such fluctuations in temperature are closely related with the near-horizon supertranslations of a Black Hole. This allows one to express the temperature-dependent corrections to the near-horizon part of the Euclidean gravity action or Free energy in terms of polynomial functionals of a near-horizon supertranslation for all orders. The leading order term turns out to be proportional to the near-horizon supertranslation charge. We also show that this same term results from a microscopic state counting of the near-horizon field with the constraint of near-horizon supertranslation symmetry imposed. The constructed near-horizon partition function provides a physically appealing intuitive description of the low-energy part of horizon physics as a sum over all possible near-horizon supertranslations. This suggests a dual description of near-horizon physics in terms of alternate variables. The implications of these results are briefly discussed.
\end{abstract}

\maketitle

\section{Introduction}
Black Holes exhibit thermodynamic properties. This provides a gateway to a microscopic theory of Black Holes and quantum gravity. However, it is not just these thermodyamic characteristics, which can be thought of as some sort of averages of different microscopic configurations, as yet unknown for Black Holes, that play an important role here. The tiny deviations from the average that are expected to show up in the low energy description of Black Holes, would contain at least partial signature of the microscopic theory. In particular, the corrections to the Bekenstein-Hawkinng entropy for Black holes have been investigated in detail for this purpose in the literature.

The Euclidean Gravity approach provides an elegant and compact way to derive several results in Black Hole Thermodynamics\cite{Gibbons:1976ue}. It has been used in the literature to study the quantum corrections to the Gravity action\cite{PhysRevLett.121.161304, PhysRevD.106.064041, AMBJORN2012127}. The natural connection to Black Hole Thermodynamics in this case makes this a convenient way to calculate the quantum corrections to Black Hole entropy also\cite{PhysRevD.51.609, PhysRevD.51.618, PhysRevD.51.R5352, PhysRevD.54.3932, Mann:1997hm, Carlip:2000nv, Govindarajan:2001ee, Page:2004xp, PhysRevD.79.044005, Solodukhin:2010pk, Sen:2012dw}. One strength of this approach lies in the way it makes direct contact with Black Hole Thermodynamcis. This makes it suitable to describe the low-energy physics of black holes also,  like the quantum corrections to Black Hole Thermodynamics\cite{PhysRevLett.133.021601} despite its limitations as regards its scope of applicability to stationary spacetimes. 
.

A notion of geometric entropy for the flat spacetime put forward and calculated in \cite{Callan:1994py} showed it to be the quantum correction to the horizon-entropy for Rindler space. For Black Holes of very large masses, the region around the black hole horizon can be described by Rindler space as a good approximation. In fact, the corrections to the entropy of a Schwarzschild Black Hole was calculated in \cite{PhysRevD.50.2700} in this manner[See also the discussion in \cite{PhysRevD.51.609} regarding this, in particular, the difference in a logarithmically divergent mass-dependent term.]. The main idea here is that the expansion of the action about the stationary point will give the quantum corrections. In this approximation, the Rindler temperature, $\frac{1}{2\pi}$, can be taken to be the temperature of the stationary Black hole. However, to get the entropy, one has to take the derivative with respect to temperature, which requires that the temperature not have some fixed value. This can be done by simply allowing the angle(Euclideanised time) around the origin point in the Euclidean spacetime to be less than $2\pi$. The price that we have to pay for this is that the event horizon will become singular and have non-zero curvature. To get the stationary point now, one has to add a constant energy density to the Euclidean gravity action at the event horizon surface. This is analogous to a spacetime in the presence of a cosmic string. One can then calculate Black Hole entropy from the Free energy of the black hole and write down a general form for the corrections to black hole entropy along with the generic form of the corrections to the Euclidean gravity itself. This form can be simply written as a polynomial of $(2\pi-\beta)$, where, $\beta$ is the inverse temperature. The temperature variation or fluctuation considered in this way remains constant along the horizon in space by construction.

 In this work, we consider the case where the fluctuation of the near-horizon temperature is taken to be a slowly-varying function of the angular coordinates for a macroscopic eternal Schwarzschild black hole. One of the main obstacle to carry this out in the Euclidean gravity approach stems from the fact that the inverse temperature is proportional to the time interval in the Euclidean action. Thus the Euclidean time period changes from point to point in space, making it difficult to determine a stationary point for the Euclidean action and also to have a physical interpretation of it. We show that this problem can be circumvented in two steps. First, we assume that the near-horizon part of the metric is invariant under supertranslations in an Effective theory of gravity. Supertranslations in this case are diffeomorphisms of the near-horizon metric in the Lorentzian signature which can be thought of as a re-labelling of the null rays, $v \to v+f(\theta, \phi)$. Such a near-horizon supertranslation can be shown to be proportional to the shift in the Killing time near the Black Hole horizon up to an approximation. This shift in the near-horizon Killing time, when continued to the Euclidean spacetime, is proportional to $\delta\beta=(2\pi-\beta)$. This allows one to interpret the change in the Euclidean time period in the near-horizon part of the action as a diffeomorphism. Thus one can allow the fluctuations of near-horizon temperature to vary on the horizon surface within the Euclidean gravity approach. However, we still need to generalise the variational problem described in \cite{PhysRevD.50.2700} to determine a stationary point of the action. For the low energy action here, we expect the variation of temperature fluctuations in the near-horizon region to be small. So we divide the near-horizon into a large number of sub-regions of much smaller size, in each of which the temperature can be taken to be constant to a good approximation. Thus one can determine a stationary point for the part of action in each such sub-region following \cite{PhysRevD.50.2700}. All of these sub-regions can then be taken together to give the stationary point in our case. Normally choosing the stationary point of a part of the action separately will not give a stationary point of the global action. In this case however, we assume the entropy and the Free energy of the Black Hole(in a vacuum) are dominated by the near-horizon contributions, so varying only the near-horizon part of the action makes sense. We have neglected the non-local terms in the action here\cite{PhysRevD.51.609} and also late time Hawking radiation.  To determine the stationary point, one has to generalise the surface energy density to be added to the Euclidean action. It will be shown here that this energy(surface integral of the surface energy density) added to the action itself turns out to be proportional to the supertranslation charge.

We face another obstacle that this task faces. When taken in totality, we would be identifying the $\tau=0$ surface with the $\tau=2\pi-\delta\beta(\theta,\phi)$ surface in the Euclideanised spacetime here. In the Euclideanisation performed originally, the Killing symmetry of the timelike vector becomes an $O(2)$ isometry in the Euclideanised space\cite{Gibbons:1976ue,PhysRevD.50.2700}. In the $\beta=constant$ case, there are two cuts at the hypersurfaces $\tau =0$ and $\tau=\beta$, which are related by the $O(2)$ symmetry. Thus they can be identified or glued together. Clearly this would not work when, $\beta\equiv\beta(\theta,\phi)$. In this case, the second cut is along the hypersurface $\tau=\beta(\theta,\phi)$, which is not related to the hypersurface $\tau =0$ by $O(2)$ symmetry.
Clearly the Killing symmetry along the vector $\partial_{\tau}$ would not suffice to glue the cuts when $\beta\equiv\beta(\theta,\phi)$.

To tackle this problem, we impose a diffeo-symmetry near $s=0$ Euclidean 'horizon' surface that shifts $\tau\to \tau+\delta\beta(\theta,\phi)$ on the Euclidean action. In practice, this is done by demanding that the integral of the extra energy density over the $s=0$ surface introduced to minimse the action be a constant. As will be shown later, it turns out that this integral is proportional to the charge corresponding to the $\tau\to \tau+\delta\beta(\theta,\phi)$ diffeo-symmetry and approximately scales as the near-horizon supertranslation charge.

However, there still remains the issue that in the approach just outlined above, one starts from a Schwarzschild Black Hole spacetime as in \cite{PhysRevD.50.2700}. As such, one is not directly analytically continuing a supertranslated Black Hole spacetime to an Euclidean one. We address this by considering the near-horizon part of the action of a Black Hole in the presence of a shockwave consisting of a thin shell of singular energy density\cite{Dray:1984ha,Blau:2015nee}. It is seen that this action indeed goes over to the near-horizon Euclidean action up to some approximation, with the singular energy density at the Euclidean 'horizon', as expected. This is not a complete resolution of the issue as its scope is limited to the level of actions. Nonetheless this strongly suggests that the temperature fluctuations should be interpreted as near-horizon supertranslations. Indeed, we can impose a diffeo-symmetry corresponding to the near-horizon supertranslation $u\to u+\delta\beta(\theta,\phi)$ by demanding that the diffeo-charge be conserved via a Lagrange multiplier. Up to some approximation, this can be analytically continued to the 'near-horizon' diffeo-charge conservation when varying the Euclidean action outlined earlier. Thus the singular energy density at the Euclidean 'horizon' required in this approach\cite{PhysRevD.50.2700} appears to have a natural interpretation in terms of shock waves inducing supertranslations.

 Once one has modified the Euclidean gravity action to accommodate temperature fluctuations, one can write down the quantum corrections to the action, in a systematic way following \cite{PhysRevD.50.2700}.These corrections turn out to be proportional to integer(+ve) powers of near-horizon supertranslation. The lowest order term, turns out to be simply proportional to the near-horizon supertranslation charge. The quantum corrections to the Black Hole entropy can be found from this in a straightforward manner. The first order term in the expression for Black Hole entropy can also be derived using a microstate counting for the near-horizon fields in a semiclassical theory, when $\delta\beta$ does not vary along the horizon[See \cite{PhysRevD.50.2700,Callan:1994py,PhysRevD.51.609}]. Since our work is an extension of the earlier method, a similar result is to be expected. We show here that this expectation is confirmed. With the form of the action now known, one can set up a path integral for it. As will be argued here, the path integral sum will be over all allowed near-horizon supertranslations.

The near-horizon supertranslations are believed to be important for Black Hole physics. It has been argued that they store information about the Black Hole\cite{Hawking:2015qqa,Hawking:2016msc}. They have been shown to play an important role in the near horizon dynamics and thermodynamics of Black Holes as well\cite{Grumiller:2019tyl, Akbar:2025wcq, PhysRevLett.120.101301,PhysRevD.101.046002,Cano:2024tcr}. Their role in construction of holographic theories for Black Hole spacetimes have also been investigated in the literature\cite{Hawking:2016sgy,Dvali:2015rea,Lust:2017gez}. The near-horizon symmetries have been studied in detail along with the fall-off conditions in the literature\cite{Donnay:2015abr, Hotta:2000gx,PhysRevD.64.124012,Donnay:2016ejv,Aggarwal:2023qwl}. The supertranslation we discuss in our paper forms only a subset of  a much larger algebra. Also we ignore the higher-order terms in the fall off conditions in considering the contribution to the Euclidean action, making our result an approximation. However, the dominant contribution to near-horizon part of this action comes from very close to the event horizon, so the approximation is a reasonable one.

The paper is organised as follows. We first introduce the approach followed in \cite{PhysRevD.50.2700} to set up the action in Euclidean gravity that has the appropriate extremal solution corresponding to a constant Black Hole temperature that is different from stationary Black Hole temperature. Then we generalise this approach to the case where the temperature fluctuations can vary in the near-horizon region. On the way, we will establish the connection between temperature fluctuations in Euclidean gravity theory and near-horizon supertranslations in the Lorentzian spacetime. Our generalised approach then leads to the result connecting the supertranslations with the Euclidean gravity action in this case as has already been highlighted in the Introduction. Next, we discuss how to set up a path integral using this action and finally conclude with a brief discussion of the physical significance of the results obtained here.

\section{Conical Singularity and Schwarzschild Black hole}
Here our discussion mainly follows the exposition given in \cite{PhysRevD.50.2700}. One starts with a Schwarzschild Black Hole of very large mass spacetime in $3+1$ dimensions. In the planar limit, the mass tends to infinity. Thus one can write down the metric in Lorentzian signature as 
\begin{equation}
\begin{split}
    ds^2 &= -\bigg(\frac{s^2}{1+\frac{s^2}{16G^2M^2}}\bigg)dT^2 + \big(1+ \frac{s^2}{16G^2M^2}\big) ds^2\\ &+ 4G^2M^2\big(1+ \frac{s^2}{16G^2M^2}\big)^2 d\Omega^2  .\label{SchwarzschildRind}
    \end{split}
\end{equation}
$G$ is the Gravitational constant and $M$ is the mass of the Schwarzschild Black Hole. $T$ and $s$ are related to the Schwarschild coordinates, $t$ and $r$ by 
\begin{equation}
\begin{split}
    T &=\frac{t}{4GM}\\
     s &=\sqrt{8GM}(r-2GM)^{\frac{1}{2}}.
    \end{split}
\end{equation}
$s$ can be viewed as the distance from the event horizon at $r=2M$. In the planar limit, \eqref{SchwarzschildRind} becomes the Rindler metric, 
\begin{equation}
    ds^2 = -s^2 dT^2 + ds^2+ (dx^2)^2+ (dx^3)^2 . \label{RindlerM}
\end{equation}
One needs to impose an infrared cutoff here to remove the divergence due to the infinite area of the Rindler horizon. A Rindler observer will find the environment thermal with the temperature $\frac{1}{2\pi}$. One can see this by Euclideanising the spacetime, with $\tau=iT$ and $\tau$ has period $\beta$, the inverse of the Black Hole temperature.

The partition function for the Rindler space is given by, $Z= Tr[e^{-\beta H_R}]$, where, $H_R$ is the Hamiltonian in the Rindler space. To calculate the entropy of the space, one defines the Free energy, $F$ by the relation, 
\begin{equation}
    Z= e^{-\beta F}  \label{FreeEnergy}
\end{equation}
and then use the thermodynamic relation, 
\begin{equation}
    S= \beta^2\frac{\partial F}{\partial\beta} .  \label{SDefn}
\end{equation}
In Euclidean gravity, 
\begin{equation}
    Z= \mathcal{N} \int Dg \prod_{i=1}^N D\chi^i \exp{\big(-S[g, \{\chi^i\}]\big)},  \label{Zdefn}
\end{equation}
where, $g$ is the metric of the Euclidean space, $\chi^i$ are the fields defined on the background spacetime and the action $S[g,\{\chi^i\}]$ is given by
\begin{equation}
    S[g,\{\chi^i\}] = S_0[g]+ S_{\chi}[g,\{\chi\}]. \label{Iparts}
    \end{equation}
    $S_0[g]$ is the Einstein-Hilbert action with the appropriate boundary term included\cite{Gibbons:1976ue}. As is well-known, this space has a conical angle deficit $(2\pi-\beta)$ leading to a conical singularity $s=0$, the point of origin in the Euclidean space. Since the origin is the event horizon of the Schwarzschild Black Hole, where the Lorentzian spacetime is regular, we remove this singularity by assuming $\beta = 2\pi$. This is the stationary point of the path integral in \eqref{Zdefn}.

    The problem with the choice of $\beta$ mentioned above is that one cannot use the formula \eqref{SDefn} to find out the entropy in this case. So we require the expression for $Z$ when $\beta\neq 2\pi$. This means we have to take into account geometries with conical singularities. 
    Such geometries will not be stationary points of the functional integral, however. To remedy that, one inserts an energy density term(analogous to a cosmic string\cite{PhysRevD.23.852,Vilenkin:1984ib}) with non-zero support only on the $s=0$ surface term in the action. This term can be thought of as the condition of fixed horizon area imposed on the variational problem.

    From the classical geometry part, one gets 
    \begin{equation}
    \int\sqrt{g}R d^4x = 2A(2\pi-\beta) , \label{EHActionSingu}
    \end{equation}
    leading to 
    \begin{equation}
        S_{EH}[g] = -\frac{(2\pi-\beta)A}{8\pi G}. \label{S_EHBeta}
    \end{equation}
Here, $A$ is the horizon area. From \eqref{FreeEnergy} and \eqref{S_EHBeta}, we get, $F= -\frac{1}{8\pi G}\frac{(2\pi-\beta)}{\beta}A$ and the entropy per unit area, $\sigma= \frac{1}{4G}$, which is just the Bekenstein-Hawking entropy.

The corrections to the Bekenstein-Hawking entropy can be found in a systematic manner starting from the following observation. Within a semiclassical theory, the first order correction digrammaticaly represents paths of first quantised particle and the higher order corrections, the Feynman diagrams. Due to the entropy formula \eqref{SDefn}, the only diagrams that would contribute to the Black Hole entropy are  those that intersect or encircle the Black Hole horizon at $s=0$. The first order correction has a divergent contribution, which can be made finite by introducing a UV cutoff\footnote[1]{Actually, there are two types of divergences, one coming from the UV cutoff and the other due to distance from the conical singularity or the brick wall cutoff. But the two energy scales can be identified\cite{PhysRevD.51.609}.}. There is a quadratic divergence, in a contribution to the Black Hole entropy that scales as its area, $A$. Thus the entropy per unit area taking this correction into account is 
\begin{equation}
    \sigma = \frac{1}{4}\Big(\frac{1}{G}+ \frac{K}{90\pi\epsilon^2}\Big), \label{S1stCorrection}
\end{equation}
where, $K$ is a constant that will be fixed by the matter fields included in the action and $\epsilon$ is the UV cutoff. The specialty of this correction term is that this keeps the action of the same form as the Einstein-Hilbert one. So one could absorb it into the bare gravitational coupling constant to define a renormalised $G_R$ in the effective action. This term can also be directly calculated in a Lorentzian spacetime setup using the brickwall method \cite{tHooft:1984kcu}, where the distance cutoff from the event horizon used is $\epsilon$.

The form of the other corrections can be determined assuming that the effective action is invariant under diffeomorphisms of the background metric $g$, the regulation of the curvature singularity at the origin and rapid falloff for Green's functions in 4D, the effective action can be written as 
\begin{equation}
    S_{Eff} = -\sum_{i=1}^{\infty} b_i(2\pi-\beta)^i .\label{S_EffCorrecn}
\end{equation}
[For details, see \cite{PhysRevD.50.2700}.] Note that $b_1= \frac{A}{8\pi G_R}$ here. \eqref{S_EffCorrecn} and \eqref{FreeEnergy} allows us to write the Free energy, $F$ as $F= -T\sum_{i=1}^{\infty} b_i(2\pi-\beta)^i$. From \eqref{SDefn}, one can now find the expression for the Black Hole entropy, 
\begin{equation}
    S= \frac{\beta A}{8\pi G_R} + \sum_{i=1}^{\infty} \{b_i+\beta(i+1)b_{i+1}\}(2\pi-\beta)^i . \label{SCorrections}
\end{equation}
It is to be noted that the expressions for both the effective action and the Black Hole entropy are now given in terms of temperature fluctuations and they reduce to the Einstein-Hilbert action and the Bekestein-Hawking entropy respectively, when fluctuations tend to zero. These terms can also be expressed as higher-derivative curvature terms as has been demonstrated in \cite{PhysRevD.52.2133,PhysRevD.50.2700}.

\section{Incorporating Temperature variations along the horizon}
So far, we have looked at the case, where the temperature of the Black Hole is constant even if it is not equal to $2\pi$. Now let us consider how to extend this formalism to include variations in this temperature. It turns out that within the approach taken here, this is possible only in the near-horizon region of the Black Hole spacetime.

\subsection{Analysing Near-horizon Temperature Fluctuations by constructing Local Grid}
The Euclidean continuations of Rindler space considered here would be such that \begin{equation}
        \frac{\lvert2\pi-\beta\rvert}{\beta}\ll 1 . \label{betafluctcond}
    \end{equation} In Euclidean gravity the temperature or it's inverse $\beta$ at the $s=0$ surface can vary along the two spatial angular coordinates. Since we want to look at the temperature fluctuations only in the near-horizon region, we shall also assume $\beta$ varies slowly\footnote[2]{For a low-energy description, this is a reasonable assumption.} in the  two spatial angular coordinates. Instead of taking the planar limit of the Schwarzschild Black hole, we shall consider here a macroscopic Black Hole. For the near-horizon region, $s\ll 4GM$ and this means that the expression for the metric given in \eqref{SchwarzschildRind} can be written as 
\begin{equation}
    ds^2 \approx -s^2 dT^2 + ds^2+ 4G^2M^2 d\Omega^2. \label{RindlerScwarzschild}
\end{equation}
Note that the near-horizon metric has the same form as the Rindler metric in $T-s$ part. We can divide the near-horizon into several sub-regions, in each of which, Rindler metric is a good approximation of the metric given by \eqref{RindlerScwarzschild}. The size of such sub-regions will be constrained by the conditions that the lengthscale,  $l_C$, inverse to the extrinsic curvature of the 3-sphere corresponding to the event horizon is much larger than it, i.e. $\lvert x^2\rvert\ll l_C$, $\lvert x^3\rvert\ll l_C$ and $\frac{s}{4GM}\ll1$.

 Because of the finite size of these Rindler patches,infrared cutoffs are no longer necessary. We assume that $\beta$ varies slowly and so it can be taken to be constant within one sub-region as a good approximation. Now we consider the part of the action in the Euclidean space for the near-horizon region only(where, $\frac{s}{4GM}\ll1$) and expand it about the stationary point as before. Proceeding as in the previous section, we can make the geometry in this sub-region a stationary point of the path integral over this sub-region(subject to appropriate boundary conditions)  by inserting an energy density term at the $s=0$ surface. As in \eqref{EHActionSingu}, for the sub-patch $\Delta_i$, this leads to the condition, $\int_{\Delta_i}\sqrt{g}R d^4x = 2A_i(2\pi-\beta_i)$, where, $\beta_i$ and $A_i$ are the time period and the horizon area respectively corresponding to the $\Delta_i$ sub-patch and the integral is over the 4-volume of $\Delta_i$. Now if we add up the actions of all these sub-patches, together they will constitute the near-horizon part of the action and the near-horizon geometry the stationary point of the path integral over the near-horizon region. Mathematically, this can be stated as
\begin{equation}
    \int_{N-H} \sqrt{g}R d^4x = \sum_i \int_{\Delta_i}\sqrt{g}R d^4x = \sum_i2A_i(2\pi-\beta_i). \label{N-HAction1}
\end{equation}

\subsubsection{Imposing the diffeosymmetry $\tau\to\tau+\delta\beta(\theta,\phi)$ in the Euclidean Near-horizon region}
At the outset, we clarify that in the Euclidean spacetime, there exists no horizon. The term 'horizon' is used only to denote a specific region in relation to its Lorentzian counterpart. As discussed in the Introduction, the identification of $\tau=0$ surface with $\tau=2\pi-\delta\beta(\theta,\phi)$ surface requires the imposition of a Diffeosymmetry in the Euclidean 'near-horizon' region. It is possible to do this by a small modification of the variational problem for the Euclidean action in a geometry with a conical singularity that has already been discussed.

The modification consists in varying the Euclidean action subject to the constraint that it obeys the diffeo-symmetry, along the vector, $\delta\beta(\theta,\phi)\partial_{\tau}$. In practice, this can  be done by imposing the constraint that, $\int_{s=0}\delta\beta(\theta,\phi)\sqrt{\gamma}d^2x =K$, where, $K$ is some constant. Here, $\int_{s=0}\delta\beta(\theta,\phi)\sqrt{\gamma}d^2x$ is the diffeo-charge[For little more details on the conservation of the diffeocharge, see part IV A of this paper.].  Note that we are assuming that the energy density is a distribution function that is non-zero only at $s=0$ as required for the variational problem in such geometries with conical singularities. The modified variational problem will be discussed in detail in the later sections.

\subsection{From Temperature Fluctuations to Supertranslations}
The Eulcideanisation procedure being carried out here is different from the standard one in that, for the near-horizon part of the action, the integral is over a time interval, $\{0,\beta_i\}$ that is different for the different sub-patches, $\Delta_i$. In the Euclidean continuation, the time period is taken to be $\beta$, thus the final state of the system being considered is identified with the initial state. Such an identification raises the question of its physical meaning.  Here this identification is being done only for the near-horizon region that is assumed to have supertranslation symmetry in the Effective Theory and we argue that in that case, the final state can be thought of as corresponding to the black hole having undergone a supertranslation.

\subsubsection{A look at the basic relations between null shifts and temperature fluctuations}
To see this, first let us define, the temperature variation, $\delta\beta = 2\pi-\beta$, and then note that, 
\begin{equation}
    \tau_i= \tau_R - \delta\beta_i, \label{STTauDiscrete}
\end{equation}
where, $\tau_R$ is the Rindler time. Now let us make the sub-regions smaller and more numerous, such that in the limit of horizon area $A_i$ corresponding to the sub-region $\Delta_i$ becoming infinitesimal, $\delta\beta$ becomes a continuous function of the two angular coordinates, i.e. $\delta\beta\equiv \delta\beta(\theta, \phi)$. In this limit, \eqref{STTauDiscrete} can be written as 
\begin{equation}
\tau = \tau_R - \delta\beta(\theta, \phi). \label{STTauContin}
\end{equation}

Now let us look at the near-horizon spacetime in the Lorentzian signature. The near-horizon metric that we have been considering can be written as 
\begin{equation}
    ds^2 \approx \frac{\rho}{2M} du^2 -2 dud\rho - 2M d\Omega^2 , \label{SchwarzschildGNC}
\end{equation}
where, $\rho= r-2M$, $r$ is the Schwarzschild radius and $\frac{\rho}{2M}\ll 1$ because we are considering the near-horizon region. $u$ is the null coordinate given by 
\begin{equation}
u = t - 2M \ln{\rho}, \label{u}
\end{equation}
for $\rho\neq0$. We note here that the expression for the near-horizon metric given by \eqref{SchwarzschildGNC} can be cast directly in the standard form in Gaussian null coordinates, 
\begin{equation}
    ds^2 = f du^2 + 2kdud\rho + 2g_{uA}dudx^A+ g_{AB}dx^Adx^B , \label{NHMetricGNC}
\end{equation}
with the required fall-off conditions, 
\begin{eqnarray}
f&=& -2\kappa\rho+ O(\rho^2), \\
k&=& 1+O(\rho^2), \\
g_{uA}&=& \rho\Theta_A+ O(\rho^2), \\
g_{AB}&=& \Omega\gamma_{AB}+\rho\lambda_{AB}+ O(\rho^2); \label{falloffcond}
\end{eqnarray}
where, $\Theta_A\equiv\Theta_A(x^A)$, $\Omega\equiv\Omega(x^A)$, $\lambda^{AB}\equiv\lambda^{AB}(u, x^A)$, $\gamma_{AB}$ is the metric of the two \cite{Donnay:2015abr}.

Now the Euclideanisation requires $t \to i\tau$ and following \eqref{STTauContin}, we can think of the final state as a transformation $\tau_R\to \tau_R - \delta\beta(\theta, \phi)$ If we go back to the Lorentzian signature spacetime now, then this suggests that, we should take $t_R - \delta\beta_L(\theta, \phi)\to i[\tau_R-\delta\beta(\theta, \phi)]$, where, $\delta\beta_L(\theta, \phi)$ has been Eulclideanised in the same way as $t$. But this would mean that the transformation $\tau_R\to \tau_R-\delta\beta(\theta,\phi)$[given by \eqref{STTauDiscrete}] in the Euclidean space corresponds to the transformation given by, 
\begin{equation}
    t\to t - \delta\beta_L(\theta,\phi), \label{ttransform}
\end{equation}
in the spacetime with Lorentzian signature. The transformation given by \eqref{ttransform} would correspond to a transformation of the null coordinate 
\begin{equation}
    u\to u-\delta\beta_L(\theta,\phi). \label{STLorentz}
\end{equation}
The transformation given by \eqref{STLorentz} is a supertranslation of the near-horizon metric given by \eqref{SchwarzschildGNC} and will leave it invariant\cite{Donnay:2015abr}.\footnote[3]{To be more precise. the transformation corresponding to the supertranslation is of the generic form, $ v\to v-\delta\beta_L(\theta,\phi)+O(\rho^3)$. We have dropped the higher order term here. }

\subsubsection{Relating the diffeo-symmetries in the near-horizon region}
We now relate the vectors corresponding to these diffeo-symmetries in the near-horizon region in both Lorentzian and Euclideanised spacetimes. For this, we focus on the $1+1$ Rindler part of the metric given by \eqref{SchwarzschildRind}, as we will not consider transformations here that mix $\theta$, $\phi$ with $t$ and $s$. This part can be expressed as the $1+1$ Minkowski metric in terms of the null coordinates, $U$ and $V$. These are related to the Rindler null coordinates $u$ and $v$ by the relations, 
\begin{equation}
    U=e^u,\ \ \ V= e^{-v}. \label{uUvV}
\end{equation}

To perform the Euclideanisation, we need to relate the shift in $u$ to the shift in Rindler time $t$ which can then be related to the shift in the Euclidean time, $\tau$ and hence to the temperature fluctuations. We will proceed in an indirect manner by first considering the shift in the Minkowski null coordinate and the related diffeo-vector. The physical reason behind this will become clear later. We will consider now a region of the spacetime that is close to the bifurcation point, but much closer to the horizon. Here this will mean $V\ll M$\footnote[5]{Note that $V$ has dimensions of length.} and $U\ll V$. This region is close to the part of the horizon, $U=0$ and $s\ll Me^{-t}$ and $t$ has large values here.

In particular, The boost Killing vector, $\partial_t$ and the basis vector $\partial_s$ in the Rindler frame are related to the basis vectors in the Minkowski reference frame($1+1$ D) by
\begin{equation}
\partial_t=\frac{\big(V+U\big)}{2}\partial_U + U\partial_T \label{tU}
\end{equation}
and
\begin{equation}
    \partial_s= 2U\partial_U-2V\partial_V, \label{sV}
\end{equation}
where, $T$ is the time in the Minkowski frame. It follows from \eqref{tU} and \eqref{sV} that $\partial_t\approx \lambda_n\partial_U$ and $\partial_s\approx -4\lambda_n\partial_V$, $\lambda_n$ being a parameter along the horizon. As far as the Euclidean basis vector $\partial_{\tau}$ is concerned, we see that it would be formally related to $\partial_t$ via the $t\to it$, Wick rotation.

Now let us take a look at the relation between the shifts in the different coordinates related to these different vectors. In the region we have been considering, we can use the rescaled coordinates, $\tilde{U}=\frac{U}{\delta}$ and $\tilde{V}=\frac{V}{\lambda_n^0}$, such that, $\delta$ and $\lambda_n^0$ are constants and $r=\frac{\delta}{\lambda_n^0}\ll1$ with, $\tilde{U}$ and $\tilde{V}$ are of the same order 1 in the region we have been considering. A shift of $U$ in this region gives $\tilde{U}\to\delta\{\tilde{U}+\tilde{f}(\theta,\phi)\}$. The relation between $u$ and $U$ leads to a shift in $u$ obeying a relation, $f_u(\theta,\phi)\approx \frac{\tilde{f}(\theta,\phi)}{\tilde{U}}$. 
Now, from the relation, $e^{-2t}=\frac{U}{V}$, it follows that, a shift in $t$ by $\delta\beta_L(\theta,\phi)$ would be related to a shift in $U$ through $\frac{2}{r}e^{-2t}\delta\beta_L(\theta,\phi)\approx \frac{\tilde{f}(\theta,\phi)}{\tilde{V}}$.($\delta\beta_L(\theta,\phi)$ is taken to be small.). The relation between $\delta\beta_L(\theta,\phi)$ and $\delta\beta(\theta,\phi)$ has already been pointed out. This makes it clear that the angular part of the shift in the coordinates scales approximately linearly with each other.

Taken together, this means that for the near-horizon region, the Euclidean time for the final state given by, $\tau= 2\pi-\delta\beta$ corresponds to a Schwarzschild Black Hole that has undergone a small supertranslation, $-\delta\beta(\theta,\phi)$ up to a scaling factor. However, this also suggests that a fluctuation of the Black Hole temperature in the near-horizon region is directly related to some supertranslation of the Black Hole itself via, $\delta\beta(\theta,\phi)\equiv-\mathcal{T}(\theta,\phi)$, where, $\mathcal{T}(\theta,\phi)$ {approximately scales as the supertranslation and $\beta$ is the inverse of near-horizon temperature. This is the first of the main results of our work.

 We assume here that the near-horizon region possesses a supertranslation symmetry. A supertranslation can also be viewed as a shift along the null rays by an amount $f(\theta,\phi)$. From \eqref{u}, we see that this would result in a shift in $t_f\to t_f+ f(\theta,\phi)$ for each $\{r=constant, \ t=t_f\}$ hypersurface. Note that such hypersurfaces span the entire near-horizon region at $t=t_f$ if we vary $r$ appropriately. The super-translation symmetry in this region is a diffeo-symmetry. It allows us to connect such $r=constant$ surfaces at time $t=t_f$ with $r=constant$ surfaces at time $t=t_f+f(\theta,\phi)$.  Because of this relation via symmetry in the near-horizon region, one can now Euclideanise $t_f\to \tau_f$, and glue the cuts in the $\tau=0$ surface to the $\tau=\tau_f+ \delta\beta(\theta,\phi)$ surface. This identification is done in two steps. First, the part of the $t=t_f$ surface in the near-horizon region is noted to be related with the part of $t=t_f+ f(\theta,\phi)$ surface there via supertranslation symmetry. Next we Euclideanise by making cuts along these hypersurfaces and glue the cuts as discussed before. }

\section{Effective Action and Black Hole Entropy}
One can find out the quantum corrections to the effective action when $\delta\beta\equiv\delta\beta(\theta,\phi)$ in two steps. First, we focus on one of the sub-regions and write  down the part of the effective action for this region following the same procedure as discussed in the previous section, when $\delta\beta$ did not vary. This step can be repeated for all the sub-regions within the near-horizon part of the spacetime. Adding these parts up, we get the near-horizon part of the effective action. It is convenient to take the continuous limit for the sub-regions at the final step. This procedure can be repeated to write down the expression for the Black Hole entropy.

Thus for the $i$-th sub-region, the part of the effective action is given by, $S_{Eff}^{\Delta_i} = -\sum_{n=1}^{\infty} b_n(2\pi-\beta_i)^n $, following the relation given by \eqref{S_EffCorrecn}, with $b_1= \frac{A}{8\pi G_R}$  as before. That is the renormalisation of the gravitational coupling constant, $G$ takes place exactly as earlier. The expression of the near-horzon part of the effective action is then found by simply adding up these contributions.
\begin{equation}
    S_{Eff}^{N-H} = \sum_iS_{Eff}^{\Delta_i}=-\sum_i\sum_{n=1}^{\infty} b_n(2\pi-\beta_i)^n. \label{SEffqcDiscrete}
\end{equation}
Now we take the continuum limit to express \eqref{SEffqcDiscrete} in integral form, 
\begin{equation}
    S_{Eff}^{N-H}\approx \int \sqrt{\gamma}d^2x \sum_{n=1}^{\infty} b'_n\{\delta\beta(\theta,\phi)\}^n\rvert_{s=0}\approx\beta F , \label{SEffectiveTVarying}
\end{equation}
where, $\gamma$ is the determinant of the metric of the $\theta-\phi$ subspace. To impose the diffeo-symmetry, we modify the procedure by varying $S^E_{E-H}(g^E)-\lambda\big(\int \sqrt{\gamma}d^2x \delta\beta(\theta,\phi)\rvert_{s=0}-K^E\big)$, where, $\lambda$ is the Lagrange multiplier and the first term within the first bracket accompanying it comes from the singular energy distribution at $s=0$ as before. The superscript $E$ denotes Euclideanised variables. Variation with respect to $g^E{\mu\nu}$ gives the same result as above and the variation with respect to $\lambda$ imposes the diffeo-symmetry along the vector $\delta\beta(\theta,\phi)\partial_{\tau}$. The Lagrange multiplier $\lambda$ can be fixed later to get the Einstein coupling between matter and energy. The near-horizon part of the effective action in Euclidean gravity for Scahwarzschild Black Hole spacetime thus turns out to be a functional of a polynomial form of the supertranslation, $-\delta\beta(\theta,\phi)$. The term linear in $\delta\beta$ here is of the form, $b'_1 \int \sqrt{\gamma}d^2x \delta\beta(\theta,\phi)\rvert_{s=0}$, which is proportional to the near- horizon supertranslation charge \cite{Donnay:2015abr}. This is the second main result of this work.

Finally, the corrections to the Black Hole entropy can be calculated in the same way as before. One can calculate it using the thermodynamic relation between the Free Energy and the entropy. However, one can also generalise \eqref{SCorrections} to directly write down the expression for Black hole entropy in this case,
\begin{equation}
\begin{split}
  S &\approx  \frac{\beta A} {8\pi G_R} \\ &+ \int_{s=0} \sqrt{\gamma}d^2x\Big(\sum_{n=1}^{\infty} \big[b_n\\&+(n+1)b_{n+1}\{2\pi-\delta\beta(\theta,\phi)\}\big]\{\delta\beta(\theta,\phi)\}^n\Big)\\\ &\rvert_{s=0}
  \end{split}
\end{equation}
Again we find that the expression of the Black Hole entropy is a polynomial functional of supertranslation, the third important result of our work. Note that, when $\delta\beta$ does not vary with $\theta$ and $\phi$, we get back the expressions for effective action and Black Hole entropy given in \cite{PhysRevD.50.2700}. It is also interesting to look at the energy density term that had to be inserted to make the solution a stationary one. Essentially, this the term that regulates the conical singularity and can be found out from that condition to be given by\cite{PhysRevD.52.2133}, 
\begin{equation}
    \sigma_E= \frac{\delta\beta(\theta,\phi)}{8\pi G},   \label{Edensity} 
\end{equation}
which has now also become position dependent. The energy term corresponding to $\sigma_E$, that has to be added to the Euclidean action, is given by, 
\begin{equation}
    S_{extra} = \frac{1}{8\pi G}\int \sqrt{\gamma}d^2x \delta\beta(\theta,\phi)\rvert_{s=0} \label{CString}
\end{equation},
which again is proportional to the supertranslation charge. 
However, there is another way of re-deriving the same energy density term, as will be seen now.

\subsection{1st order correction from microstate counting}
To determine the stationary point, we start by imposing a constraint on the extremization problem that near-horizon part of the action should possess a supertranslation diffeo-symmetry in the Lorentizan space. This is done by adding the symmetry-constraint via a Lagrange multiplier to the action. We expect the dominant contribution to Black Hole entropy or the Free energy, $F$ to come from the near-horizon region part of the action only when continued to Euclidean space.

The diffeo-constraint is then,
\begin{equation}
    \delta_{\xi}S_{NH} =0,
\end{equation}
where, $\xi\equiv \big(\mathcal{T}(\theta,\phi), 0,0,0)$ is the vector characterising the super-translation diffeo-symmetry of the action, $S_{NH}$. This leads to the continuity equation,
\begin{equation}
    \int\nabla_{\mu} T^{\mu\xi} = 0 . \label{ContinuityEqn}
\end{equation}
$T_{\mu\nu}$ is the energy-momentum tensor in the near-horizon region. 
From \eqref{ContinuityEqn}, we get the conservation equation for the diffeo-charge,
\begin{equation}
    \int_{NH}\mathcal{T}(\theta,\phi)T^{vv} \sqrt{h}d^3x = C, \label{CConserv}
\end{equation}
where, $C$ is some constant. In the near-horizon region, we can replace $T^{vv}$ by $T^{tt}$, where, $t$ is the Killing time as a good approximation. Then charge conservation equation leads to 
\begin{equation}
     \int_{NH}\mathcal{T}(\theta,\phi) T^{tt} \sqrt{h} d^3x \approx C. \label{CConservSemiclass}
\end{equation}
The variational problem then becomes, 
\begin{equation}
    \delta\big(S_{NH}-\lambda\int_{NH}\mathcal{T}(\theta,\phi) T^{tt} \sqrt{h} d^3x\big)=0. \label{VarLagrange1}
\end{equation}
We know that to find the stationary point of the Euclidean action, the energy density, $T{tt}$ in the second term on the l.h.s. of \eqref{VarLagrange1} has to be a surface energy density at the event horizon. This fixes $\lambda=\frac{1}{8\pi G}$ and gives the same energy density as in \eqref{Edensity}.

A problem arises however if we consider quantisation. The near-horizon fields do have a dominant contribution to the energy density, $\langle T_{tt}\rangle$, that exist on the horizon, but it is divergent. Here $\langle T^{tt}\rangle$ is the semiclassical expectation value in the near-horizon region. 
To be concrete, we consider this here for a free scalar field put in the Black hole background and it's energy expectation value can found from a semiclassical calculation using the brickwall method\cite{tHooft:1984kcu,Frolov:1998vs,PhysRevD.50.2700,PhysRevD.78.024003}. The wave modes for the scalar field are determined using a reflected boundary condition and a thermal environment at the Black Hole temperature. $\langle T^{tt}\rangle$ in this case is thus the thermal expectation of the energy of the scalar field in the near-horizon region. This is given by, 
\begin{equation}
\bar{E} = -\frac{\partial}{\partial\beta}\ln{Z_{Sc}} = \frac{KA_H}{(2\pi)^3\epsilon^2}, \label{EbarNH}
\end{equation}
where, $Z_{Sc}$ is the corresponding semiclassical partition function, $A_H$, the horizon area and $K$ is a constant. $\epsilon$ is the brick wall cutoff put to regulate the divergence. One thus finds that the dominant contribution to the near-horizon energy density is a surface density, which moreover is singular at the event horizon. To regulate it, we put a counter-term with opposite sign in the action and renormalise the gravitational coupling constant by the relation, $\frac{1}{G_R} = \frac{1}{G}+\frac{KA_H}{(2\pi)^3\epsilon^2}$. $G$ is now viewed as the bare coupling constant. Thus effective action has a term linear in $\delta\beta$ given by, $\frac{1}{8\pi G_R}\int \sqrt{\gamma}d^2x \delta\beta(\theta,\phi)\rvert_{s=0}$, the same as before.\footnote[4]{There is a subtlety here. The constraint of the diffeocharge imposed for the Lorentzian theory is continued to the Euclidean theory by contour deformation. Putting \eqref{EbarNH} in the expression for the diffeocharge, however, makes it divergent when $\epsilon\to0$. This creates a problem in the contour deformation. However, for $\epsilon\neq0$, this problem goes away. Thus, strictly speaking, only the renormalised theory can be continued to the Euclidean spacetime. We thus find from \eqref{CString} that the extra term added to the action in \eqref{N-HAction1} is proportional to the diffeo-charge. This suggests that varying the action in equn (15) at the semiclassical level is effectively imposing the diffeo-symmetry via a Lagrange multiplier term for a constant diffeo-charge. Of course, for this, one has to add a constant term to the action which would not change the equations of motion for $g_{\mu\nu}$s.}


\subsection{Lorentzian action with a massless shockwave and Euclidean geometry with a conical defect}
Now we take a very brief look at the level of actions(their near-horizon parts) at a close connection between massless shockwave near the Schwarzschild Black Hole horizon in Lorentzian spacetime and the near-'horizon' part of the action for the Euclideanised geometry with a conical defect and a singular energy density at the Euclidean 'horizon'.We find they are closely connected providing further evidence that the temperature fluctuations in the Euclidean context are related to near-horizon supertranslations in the Lorentzian spacetime.

We start by considering a massless shockwave in the near- horizon region of the Schwarzschild Black Hole.  As earlier, we can divide this region into smaller grids, where, each smaller segment is the local neighbourhood of given values of $\theta$ and $\phi$. We can use a Minkowski metric and a Rindler metric to describe the spacetime with the characteristics specified($V\ll M, U\ll V$) in the part III.B.2 in such a local neighbourhood.  In each of these segments, one of the Minkowski null coordinate undergoes a  shift at a given null hypersurface or more precisely a shift on the part of the null hypersurface that falls within the region we are considering here. The null hypersurface, where this shift occurs is the one, where, a shockwave is said to exist. This is the well-known Dray-t'Hooft shell inducing a supertranslation in one of the Minkowski null coordinates\cite{Dray:1984ha}. The shockwave has a singular energy distribution which is constant. In our case, we are applying it only locally in each of the segments. Now for each segment, the shift is different, making it $(\theta,\phi)$-dependent on the entire near-horizon region of the Black Hole, i.e. of the form, $U\to U+f(\theta,\phi)$. To have a smooth solution across these segments, we vary the action subject to the diffeo-symmetry along the vector, $f(\theta,\phi)\partial_U$. This translates to a variational problem, $S=S_{E-H}+\lambda\bigg(\int T_{UU}d\Sigma - K\bigg)$ for a segment, where, $\lambda$ is the Lagrange parameter.[$T_{\mu\nu}\neq 0$, only for $T_UU$ in the $(\partial_U, \partial_V,\partial_{\theta},\partial_{\phi})$ basis.] For the singular energy density for the shockwaves, this becomes $S=S_{E-H}+ \lambda\big(T_{UU}\Delta A-K\big)$. Adding up the contributions for all the segments and varying it, implies, 

\begin{equation}
\int_{\Sigma_U}f(\theta,\phi)= constant.\label{STChargeJunc} 
\end{equation}

The energy momentum tensor in each of these segments can now be found in the Rindler frame. From the \eqref{tU} and \eqref{sV} and the approximations used in part, III.B.2, one gets $T{\mu\nu}\partial_t^{\mu}\partial_t^{\nu}\approx V^2T_{UU}$ and $T{\mu\nu}\partial_s^{\mu}\partial_s^{\nu}\approx 0$ and $T{\mu\nu}\partial_t^{\mu}\partial_s^{\nu}\approx 0$. \footnote[7]It is to be noted here, that apart form the energy density, the other components of the energy momentum tensor in this $t-s$ basis are only approximately zero.This means that in this segment, we can write the action to vary as 
$S_{\Delta}\approx S_{\Delta}^{E-H}+\lambda\int\big(T_{tt}-K'\big) d^4x$. If we now Euclideanise it by going from $t\to \tau$, we get the action to be $S^E_{\Delta}\approx S^{E-H}_{E\ \Delta}+\lambda \big(\int T_{\tau\tau}d^4x- K'^E\big)$. $T_{UU}$ for the shockwave is a singular energy density at some null hypersurface. Thus one gets the singular energy density on a small portion of a $S^2$ surface(because of spherical symmetry in this case) for a range of the null coordinate and it becomes a small portion of the $S^2$ on the horizon as one approaches to the horizon. This is the $s=0$  hypersurface. Thus, one expects the Euclideanised action in such segments to be $S^E_{\Delta}\approx S^{E-H}_{E\ \Delta}+\lambda\Big(\int_{s=0}T^E_{tt}\vert_{\Delta}\sqrt{\gamma} d^2x - K'^E\Big)$. Adding up the contributions from these segments, one gets, the familiar 'near-horizon' action, $S^E=S^E_{E-H}+\lambda\big(\int_{s=0}\delta\beta(\theta,\phi)\sqrt{\gamma}d^x- K^E\big)$ to vary. Sketchy as this argument is, it still suggests the central claim, the near-horizon temperature fluctuations are closely related to near-horizon supertranslations.


\subsection{The near-horizon partition function}
We can write down the partition function by putting the expression for $S_{NH}$ given by \eqref{SEffectiveTVarying} in \eqref{Zdefn}. Normally, carrying out the gravitational path integral is a challenging task. The gravitational action need not be positive definite, one has to gauge away the diffeosymmetries while the sum is over different geometries\cite{Gibbons:1978ac,PhysRevD.18.1747,Mazur:1989by,Banihashemi:2024weu,Banihashemi:2022jys}. In this case however, we are looking at only the near-horizon part of the effective Euclidean gravity action. Furthermore, it can be expressed as a series, whose terms are polynomial functionals of the temperature fluctuation, $\delta\beta(\theta,\phi)$. This suggests a rather simple prescription for the near-horizon part of the path integral at low energies.

There are no restrictions on $\delta\beta$ as long as it is analytic and small as required by \eqref{betafluctcond}. So the path integral is to be carried out by summing over all allowed $\delta\beta(\theta,\phi)$. We have already argued about how the $\delta\beta(\theta,\phi)$ can be thought of as a near-horizon supertranslation. So we can view the near-horizon partition function as a sum over all the allowed near-horizon supertranslations.

\section{Discussion}:
Near-horizon supertranslations are now believed to play an important role in Black Hole physics, particularly regarding the information stored in Black Holes. This raises the interesting question, how does the Black Hole entropy depend on supertranslations. This work attempts to throw some light on this issue by looking at the quantum corrections of the Black Hole entropy and finding it can be expressed in a simple form in terms of near-horizon supertranslations. In fact, the low-energy corrections to the near-horizon effective action can be expressed in terms of such a supertranslation. Non-local correction terms has not been considered here however[See \cite{PhysRevD.51.609,Fursaev:1994in}]. For a low-energy theory, we expect such terms to be suppressed. In fact, the terms we have put in are the ones we obtained by means of some approximations. So there would be other terms in the effective action not discussed here.

The direct relation of the supertranslations with the variation of the Black Hole temperature on the Black Hole horizon suggests a Thermodynamic-Statistical Mechanical interpretation for near-horizon supertranslations. This is further backed up by the appearance of supertranslation in the expression for near-horizon Free energy. It tells us that depending on the supertranslation,the corrections to the BH entropy is different. This suggests a direct connection of the near-horizon supertranslation with information stored by Black Holes.

We have given here a form for the effective action for the near-horzion region and also outlined how one can construct a Euclidean path integral from that. This might be useful in considering a scattering of wavepackets by the Black Hole. The form of the low-energy quantum corrections to the scattering amplitude might be determined or at least constrained by such an analysis. Our analysis also suggest how to generalise the approach put forward in \cite{PhysRevD.50.2700,Callan:1994py,PhysRevD.51.609} in the case when the horizon temperature variation becomes a function of spatial coordinates. It would be interesting to see whether this is connected to some kind of microscopic fluctuation/violation of the Zeroeth law of Black Hole Thermodynamics.

Our result appears to hint at a dual description of the near-horizon physics in terms of alternate variables, e.g. near-horizon supertranslations. Already at the classical level, such a description exists, e.g. the membrane paradigme\cite{Thorne:1986iy}. It would be interesting to explore whether our results form part of the outlines of a similar theory at the quantum level. In fact, the possibility of describing the near-horizon physics using the near-horizon symmetries has been explored in the literature\cite{PhysRevLett.120.101301,PhysRevD.101.046002,Penna:2015gza}.

Finally, we have looked here only at the simplest case, the Schwarzschild Black Hole. It would be interesting to explore the relation between supertranslations and temperature fluctuations in the case of Kerr Black Hole. The angular dependence of the metric components themselves is likely to make the analysis complicated there.

\bibliographystyle{unsrt}
\bibliography{references}

\end{document}